\documentclass[12pt]{article}
\usepackage[english,slovene]{babel}
\usepackage[utf8]{inputenc}
\usepackage{epsfig,amssymb,latexsym,color,amsmath,pifont}
\newcommand{\n}{\noindent} 
\newcommand{\e}{\varepsilon}
\newcommand{\om}{\omega}

\newcommand{\x}{\mathbf{x}}
\newcommand{\T}{\mathbf{T}}
\newcommand{\Z}{\mathbb{Z}}

\begin{document}

\title{\textcolor{blue}{\textbf{Ergodic theory and visualization. II. Harmonic mesochronic plots visualize (quasi)periodic sets}}}
\author{Zoran Levnaji\'c${}^{1,2}$, Igor Mezi\'c${}^{2}$} 
\date{}
\maketitle

\begin{center}  \begin{minipage}{6.2in}
\begin{flushleft}
\begin{small} ${}^{1}$\textit{Faculty of Information Studies in Novo mesto, 8000 Novo mesto, Slovenia}  \end{small} \\
\begin{small} ${}^{2}$\textit{Department of Mechanical Engineering, University of California Santa Barbara, Santa Barbara, CA 93106, USA} \end{small}
\end{flushleft}

We present a new method of analysis of measure-preserving dynamical systems, based on frequency analysis and ergodic theory, which extends our earlier work~\cite{vis1}. Our method employs the novel concept of harmonic time average~\cite{mb}, and is realized as a computational algorithms for visualization of periodic and quasi-periodic sets or arbitrary periodicity in the phase space. Besides identifying all periodic sets, our method is useful in detecting chaotic phase space regions with a good precision. The range of method's applicability is illustrated using well-known Chirikov standard map, while its full potential is presented by studying higher-dimensional measure-preserving systems, in particular Froeschl\'e map and extended standard map.
\end{minipage}  \end{center}

\n \textbf{Keywords}: dynamical systems, measure-preserving maps, ergodic theory, computational visualization, periodic sets, frequency analysis, chaotic dynamics \\[0.3cm]

\n \textbf{Lead paragraph}  Equations modeling most of the real-world systems are usually impossible to solve. For this reason, a researcher must resort to computational methods approximating the actual solution, or identifying specific details of interest without solving the system. Using newly proposed concept of harmonic time average, we designed a novel method of analyzing dynamical systems. Our method relies on two well established methodologies: frequency analysis, a tool for decomposing the motion into separate easily tractable components, and ergodic theory, a powerful mathematical framework for statistical analysis of motion. For illustrating the implementation of our method, we employ Chirikov standard map, a widely studied and well understood chaotic system. As we show, our method is able to detect and graphically visualize any periodic set within the dynamical phase space, that is, a set composed of several disjoint parts that are periodically visited. Our method is applicable to systems of arbitrary dimensionality and works regardless of their integrability. In fact, we foresee the main application exactly in the domain of high-dimensional system.

%% -----------------------------------------------------------------------------------------------------------------------------------------------------------------------------------
%% -----------------------------------------------------------------------------------------------------------------------------------------------------------------------------------
%% -----------------------------------------------------------------------------------------------------------------------------------------------------------------------------------

\section{Introduction} \label{Introduction}

Many problems in modern science and technology can be formalized via dynamical systems, which are a powerful framework for mathematical and computational analysis~\cite{dynsys-books,meissmaps}. Yet, for a model of a given real-world process to be even remotely realistic, it inevitably ends up being formulated as a high-dimensional system of equations~\cite{igorpnas}. This calls for the development and improvement of computational methods aimed at analyzing dynamical systems, with or without solving them explicitly. Over the past decades, parallel to the fascinating growth of the computational power, also grew the number of analytic tools for complex dynamical systems~\cite{froeschle-science}, including applications as diverse as diagnostics of oil spill movement~\cite{sophie-science}.

When choosing from the spectrum of the available methods, except taking into account the nature of the problem, one also considers the aims that the study is to achieve~\cite{jones}. The choice of method is often the decisive factor for the overall efficiency of the conducted analysis. Dynamical systems are still typically analyzed via direct integration of trajectories, taking into account the ratio between precision and numerical cost. Some methods are based on statistical analysis of the dynamics, such as return time plots~\cite{thiere} and exit time plots~\cite{meissexit}. Others revolve around looking for invariant sets and measures~\cite{dellnitz} or stable and unstable manifolds~\cite{krauskopf}, which can be computed via fat trajectories~\cite{henderson-fat}.

Of particular interest if often to graphically visualize certain aspects of the dynamics, for example the phase space structure. Visualization methods usually work as algorithms for dividing the phase space into subsets according to a prescribed property. In our previous publication~\cite{vis1}, we presented a method for visualization of invariant sets, based on the ergodic partition theory~\cite{mezic}. By computing the dynamical time averages of chosen functions, we constructed Mesochronic Plots (MP) allowing to graphically identify all invariant sets and thus obtain the invariant phase space partition. Our method relied on applying the \textit{ergodic theory}~\cite{ergodic-books}, whose power in the statistical modeling of dynamical systems can be seen in both theory~\cite{dalessandro} and experiment~\cite{soti}. In this paper, we extend our earlier work, and present a novel visualization method, able to detect periodic phase space subsets or any periodicity (or frequency).

In fact, \textit{frequency analysis}, except being widely used for data analysis~\cite{freqanal-books,hou}, is also a powerful tool for studying motion. It is applied primarily by decomposing the dynamics into harmonic components, which is elegantly done via spectral decomposition~\cite{igordecomp}. The earliest implementation involves estimating the system's fundamental frequencies~\cite{laskar}, which eventually opened new ways of quantifying chaos. Another approach relying on wavelets allows the detection of resonance trappings and transitions~\cite{chandre}, specifically useful for examination of weakly chaotic orbits. Frequency analysis is shown to provide insights into the difference between noisy flows and the colored noise~\cite{jun}. And of course, this framework has interesting applications other than dynamical systems, notably in neuroscience~\cite{canolty} and in the study of sound and music~\cite{oppenheim}.

There are several ways to ``marry'' the frequency analysis with the ergodic theory~\cite{petersen}. One of them is based on defining the \textit{harmonic time average} (HTA), as a generalization of the usual functional time average known from ergodic theory~\cite{mb}. It can be shown, that the constancy of the absolute value of HTA (HTA is complex-valued), is related to the invariance of the underlying phase space portion, in a way analogous to the usual time average ~\cite{mauroy,budisic2}. In this paper we present a new computational visualization method, constructed by computing HTA over the grid of initial phase space points. Such 2D plot, computed for a prescribed frequency $\om$ (or periodicity $p=\frac{1}{\om}$), we call Mesochronic Harmonic Plot (MHP), in analogy with MP from~\cite{vis1}. Periodic set of periodicity $p$ are revealed as joint level sets of the absolute value of HTA. When multiple linearly independent functions are considered, one obtains Mesochronic Harmonic Scatter Plot (MHSP), in analogy with MSP from~\cite{vis1}. This can be used to identify the periodic partition of the phase space (meaning, the partitioning of the phase space into a union of disjoint periodic sets of prescribed periodicity). Such partition can then be graphically visualized by coloring the separate subsets differently. The precision of the obtained partition depends on the number of linearly independent functions considered. As we also show in what follows, our method is closely related with the spectral decomposition of Koopman operator, while it is complementary to the analysis via classical Fourier transform. Also, note that out method is fully supported for arbitrary measure-preserving system, regardless of its integrability (which contrasts many of the above mentioned methods).

Upon exposing the mathematical details of our method, we first present a simple implementation via Chirikov standard map~\cite{chirikov}, whose dynamical properties have been widely studied and are well understood. Once the working and the applicability range of our method is clear, we demonstrate its power by analyzing higher-dimensional dynamical systems, in particular Froeschl\'e map~\cite{froeschle} and extended standard map~\cite{esm}. In fact, the full potential of our method lies in the study of high-dimensional dynamical systems, which can be most easily done by examining MHPs of lower-dimensional surfaces/planes.

%% -----------------------------------------------------------------------------------------------------------------------------------------------------------------------------------
%% -----------------------------------------------------------------------------------------------------------------------------------------------------------------------------------
%% -----------------------------------------------------------------------------------------------------------------------------------------------------------------------------------

\section{The Visualization Method} \label{The Visualization Method}

We begin by presenting the mathematical basis of our method. Since our interest here is exposing the method's applicability, we will skip the rigorous proofs and refer the reader to~\cite{mb,ergodic-books}. 

Consider a measure-preserving map $\T$ on a compact metric phase space $A$ evolving in discrete time $t$:
\begin{equation}  \x_{t+1} = \T \x_t, \;  t \in \Z, \ \ \x_t \in A. \label{eq-map} \end{equation}
An \textit{invariant set} $B \subset A$ for the dynamics $\T$, is a set such that each trajectory that starts in it stays in it forever~\cite{dynsys-books}:
\[ \x  \in  B \;\; \Leftrightarrow \;\; \T^t \x  \in B \; \;\; \forall t \in \Z . \]
We are here interested in the specific class of invariant sets called \textit{periodic sets}. A period-$p$ set $B_p$ is an invariant set composed of $p$ disjoint subsets; $B_p = \cup_{k=1}^p B_{p}^{(k)}$, with the following property:
\begin{equation}  \x  \in  B_p^{(k)} \; \Rightarrow \; \T \x  \in B_p^{(k+1)}, \;  \T^2 \x  \in B_p^{(k+2)}, \hdots \; \T^p \x  \in B_p^{(k+p)} \equiv B_p^{(k)}.  \label{eq-pe-set} \end{equation}
Therefore, a trajectory starting in $B_p^{(k)}$ visits the entire cycle $B_p^{(k+1)}, B_p^{(k+2)} \hdots$ visiting each subset once, before coming back to $B_p^{(k)}$ after $p$ iterations. Periodic set can be seen as a generalization of periodic orbit, while period-1 set is of course the usual invariant set. 

In our last paper~\cite{vis1}, we proposed a method based on ergodic theory aimed at identifying the invariant sets. Below we present the extension of our method, originally introduced in~\cite{mb,budisic2}, for identifying the periodic sets of predetermined periodicity $p$ within the phase space of any measure-preserving map Eq.\ref{eq-map}. 

The \textit{time average} $f^*$ of a function $f$ under the map Eq.\ref{eq-map} is defined as:
\begin{equation} f^* (\x) = \lim_{t \rightarrow \infty} \; \frac{1}{t}  \sum_{k=0}^{t-1}  f (\T^k \x) ,  \label{eq-ta-def} \end{equation}
for almost every (a.e.) $\x \in A$ and for every $f \in L^1(A)$~\cite{ergodic-books}. 
%A map $\T$ is termed \textit{ergodic}, if there exists a measure $\mu$ such that the time average $f^*$ of any $f \in L^1(A)$ is equal to its space average:
%\begin{equation} f^* = \frac{1}{\mu(A)} \int_{A} f d \mu , \end{equation}
%a.e. in $A$. Each $f^*$ is an \textit{invariant function}, since $f^* (\x) = f^*(\T^t \x) \; \forall t$. 
The algorithm presented~\cite{vis1} relies on computation of the time averages and operates by decomposing the phase space into the ergodic partition, which is identified as partitioning into a union of invariant sets. The extension hereby presented employs the novel concept of harmonic time average (HTA).

Consider the map Eq.\ref{eq-map} and a real-valued function $f \in L^2(A)$. For a given frequency $\om \in [0,\frac{1}{2}]$ and $\x \in A$, we define the HTA $f_\om^*$ as:
\begin{equation} f_\om^* (\x) = \lim_{t \rightarrow \infty} \; \frac{1}{t} \sum_{k=0}^{t-1} e^{i 2 \pi k \om} f (\T^k \x) .  \label{eq-hta-def} \end{equation}
HTA generalizes the concept of time average Eq.\ref{eq-ta-def}, since for $\omega=0$ we have $f_{\omega}^* = f^*$. By the Ergodic Theorem, the limit Eq.\ref{eq-hta-def} exists a.e. in $A$ for any measure-preserving map $\T$ and function $f$~\cite{ergodic-books}. In contrast to Eq.\ref{eq-ta-def}, HTA $f_\om^*$ is complex-valued, and is not an invariant function, since:
\[  f_\om^* (\T^t \x)  = e^{- i 2 \pi \omega t} f_\om^* (\x) . \]
However, the absolute value (radius) of $f_\omega^*$ is an invariant function:
\begin{equation}  |f_{\omega}^{*} (\T^t \x)|  =  |f_{\omega}^{*} (\x)| . \end{equation}
This means that, in equivalence with the usual time average Eq.\ref{eq-ta-def}, constancy of $|f_{\omega}^{*}|$ over some subset of the phase space indicates the invariance of that subsets. In the reminded of the this paper we will be mostly dealing with $|f_{\omega}^{*}|$, so we adopt the notation $h_\om = |f_{\omega}^{*}|$.

The key result enabling our visualization method, is that the HTA for a point belonging to period-$p$ set, can be non-zero only if the HTA is computed for the periodicity $p$ (frequency $\om=\frac{1}{p}$). In all other cases the HTA eventually averages out to zero. To illustrate this, we note that the expression averaged in Eq.\ref{eq-hta-def} depends on the discrete time $k$ through the radial part $f(\T^k \x)$ and the phase part $e^{i 2 \pi k \omega}$. The latter directly relates to the selected frequency $\omega$. If the selected frequency (periodicity) coincides with that of a given set $B_p$, the sampling of complex values will exhibit a periodic pattern, allowing for HTA to converge to a non-zero (complex) value. Otherwise, the sampled complex values will be randomized in phase, and regardless of the choice of $f$ will always average out to zero. 

To present this key results more formally, we begin by considering the \textit{shift map} $\Theta_\om$, defined as the mapping of the circle $S^1 \equiv [0, 2\pi[$ onto itself:
\[ \Theta_{\omega} \theta = \theta + 2 \pi \omega , \;\; \omega \in [0,\frac{1}{2}] \; . \]  
Under the shift map with frequency $\om$ every subset of the circle is periodic with periodicity $p=\frac{1}{\om}$. Now, consider a map $\T$ on phase space $A$, which admits a periodic set $B_p$. The shift map $\Theta_\om$ is called the \textit{factor map} to $\T$ on $B_p$, if there exists a measure-preserving homeomorphism $F: A \rightarrow S^1$ such that
\[ (F \circ \T) (\x) = (\Theta_\om \circ F) (\x) \;\; \forall \x \in B_p \; . \]
This implies that on $B_p$, the map $\T$ is topologically equivalent to $\Theta_\om$, which means that the set $B_p$ is a periodic set (or a union of periodic sets) with the period $p=\frac{1}{\om}$. The following two theorems hold:
\paragraph{Theorem 1} Given a map $\T$ and $f \in L^2(A)$, if $h_\om$ is non-zero on some set $B_p \subset A$, then there exists a factor map for $\T$ on $B_p$ given by the shift map with the frequency $\om=\frac{1}{p}$.
\paragraph{Theorem 2} If a map $\T$ admits a shift map with the frequency $\om$ as a factor map on some $B_p \subset A$, then there exists an $f \in L^2(A)$ such that $h_\om$ is non-zero on $B_p$.\\[0.15cm]

As we show in the following Sections, our method is implemented by computing the $h_\om$ for a selection (grid) of initial phase space points. This computation divides the phase space into two regions: the region composed of periodic set resonating at the selected frequency (HTA non-zero), and the remaining non-resonating region (HTA zero). Note that the final value of $h_\om$ also depends on the choice of $f$, and in fact, for some choices of $f$, $h_\om$ can average out to zero despite the correct frequency. Hence, in order to visualize all periodic sets in their entirety, one needs to consider for each frequency more linearly independent functions $f$.

A shift map with the frequency $\om$ is also a shift map with the frequencies that are integer multiples of $\omega$. Therefore, a HTA of frequency $\omega=\frac{1}{p}$ reveals all periodic sets with  periodicities that are integer multiples of $p$. The most meaningful choices for the periodicity are thus mutually prime values. We call \textit{period-$n$ partition} the division (partitioning) of phase space into a union of disjoint period-2 sets.

HTA analysis is related to the eigen-decomposition of the Koopman operator $U$, defined for the map Eq.\ref{eq-map} via composition:
\[ (U f) \x = (f \circ \T) \x \; . \]
$U$ acts as an evolution operator in the functional space as $(U f) \x = f (\T \x)$, enabling an alternative representation of the dynamics. It follows that the $f_\omega^*$ is an eigenfunction of $U$ with the eigenvalue $e^{-i 2 \pi \omega}$:
\[ U f_\omega^* = e^{-i 2 \pi \omega} f_\omega^* . \]
This is to say that the HTA provide a way to decompose the Koopman operator. Of course, the usual time average Eq.\ref{eq-ta-def} is the eigenfunction of $U$ with the eigenvalue 1.

We also here relate the HTA with the frequency analysis of dynamical systems, by observing the difference with the Fourier transform. Frequency decomposition of a selected trajectory computed via Fourier transform reveals all the frequencies that compose that trajectory. In contrast, HTA is typically computed for a large set of trajectories (initial points), yielding the set of resonating and non-resonating values, but only for the single chosen frequency.

%% -----------------------------------------------------------------------------------------------------------------------------------------------------------------------------------
%% -----------------------------------------------------------------------------------------------------------------------------------------------------------------------------------
%% -----------------------------------------------------------------------------------------------------------------------------------------------------------------------------------

\section{Single Function Plots} \label{Single Function Plots}

In this Section we present the implementation our method relying two-dimensional (2D) Chirikov standard map~\cite{chirikov}. Its dynamical properties are well known and still widely studied~\cite{ja2}. It is a measure-preserving (symplectic) map, mapping the square $[0,1] \times [0,1] \; \equiv [0,1]^2$ onto itself:
\begin{equation} \begin{array}{lllc}
  x' &=& x + y  + \e \sin (2 \pi x)  \;\;\;  &[mod \; 1]  \\
  y' &=& y + \e \sin (2 \pi x)  \;\;\;  &[mod \; 1]
\end{array} \label{eq-sm}  \end{equation}
(the usual standard map's parameter $k$ is $k=2\pi \e$).  The map possesses a very rich structure of periodic sets changing with the parameter $\e$. We set up an orthogonal grid of $D \times D$ points on the square $[0,1]^2$ using $D=800$. We then compute the HTA for each grid point after $T$ iterations. Thus we obtain a grid of $D \times D$ values of $h_\om$, which is our Mesochronic Harmonic Plot (MHP), visualized by assigning a color-value to each grid-point. We employ on the log-scale since the obtained values range over six decades. The number $T$ is set in accordance with the convergence properties (we mostly use $T = 30000$). For simplicity, we mostly employ a single Fourier function $f(x,y)=\sin (\pi x)$. 

\begin{figure}[!ht]  \begin{center} 
       \includegraphics[width=0.7\textwidth]{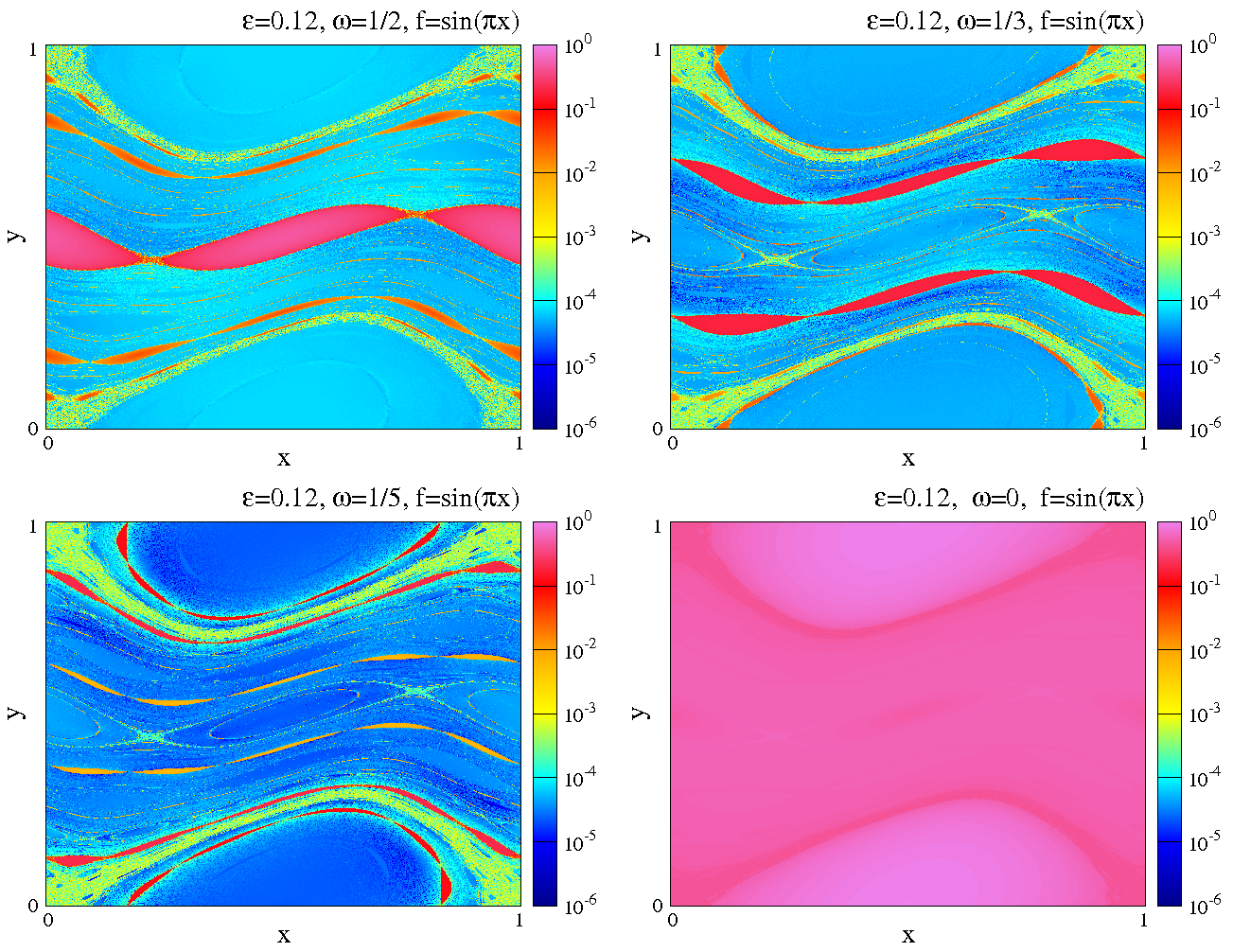}
  \caption{Single-function plots of $h_\omega$ for $\e=0.12$ and various frequencies $\omega$.} \label{fig-vis2-sin1pix-e012-variousom}
\end{center}  \end{figure}

In Fig.\ref{fig-vis2-sin1pix-e012-variousom} we show four plots for $\e=0.12$ and different frequency values. The known periodic sets of all periodicities are correctly visualized. In the $\om=\frac{1}{2}$ ($p=2$) plot, the central period-2 set is clearly pronounced. As expected, all sets of higher even periodicity ($4, 6 \hdots$) are also pronounced (standard map Eq.\ref{eq-sm} is on torus). In the plot for $\om=\frac{1}{3}$ ($p=3$), two largest period-3 sets centered around period-3 orbits are visible, along with period-6 sets and period-12 sets. 
Similarly, four period-5 sets and a single period-10 set can be seen in $\om=\frac{1}{5}$ plot. Many periodic sets of higher periodicities can be also recognized (due to limited resolution) as horizontal curves. These periodic sets resonate with the selected frequency, and have the $h_\om$-values in range of $O(10^{-1})$. In contrast, non-resonating periodic sets in all three plots have $h_\om$-values in range of  $O(10^{-5})$, which clearly distinguishes them from the resonating periodic sets. Interestingly, the chaotic zone around the hyperbolic fixed point $(0,0)$ can be also recognized in these three plots, since its $h_\om$-values are between the resonating and the non-resonating value range, in range of  $O(10^{-3})$. As stated in the two Theorems in the previous Section, $f_\om^*$ for all non-resonating points converge to zero. However, the convergence for the non-resonating periodic sets is much faster than for chaotic zone. In fact, the chaotic zone weakly resonates at all frequencies, which in principle allows its visualization. We further develop this argument later. The last plot in Fig.\ref{fig-vis2-sin1pix-e012-variousom} is done for $\om=0$, equivalent to $\om=1$. The entire phase space, and therefore all the periodic sets resonate at this frequency. This is the only frequency where the period-1 sets resonate (around fixed point at $(\frac{1}{2},0)$). For $\om=0$, HTA reduces to the time average Eq.\ref{eq-ta-def} with values in the range of $O(10^{-1})$, thus visualizing the usual invariant sets, including the chaotic zone. While useful in detecting the invariant sets, the time average Eq.\ref{eq-ta-def} cannot reveal their periodicity, since all periods are multiples of 1. All periodic sets visualized in Fig.\ref{fig-vis2-sin1pix-e012-variousom} have internal structure involving smaller periodic sets with diverse periodicities, which are not visible due to resolution limitations.

\begin{figure}[!ht]  \begin{center} 
       \includegraphics[width=0.7\textwidth]{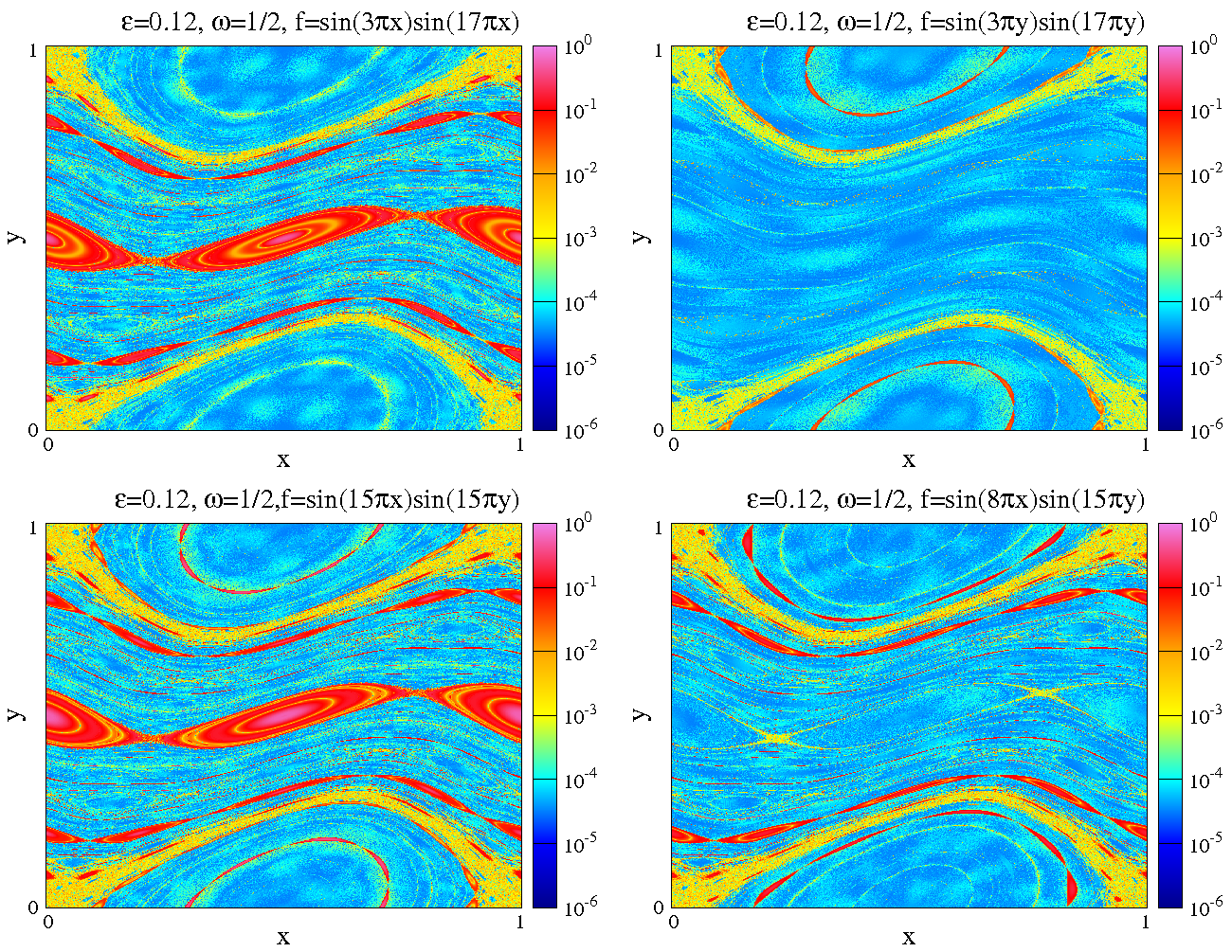}
  \caption{Single-function plots of $h_\omega$ for $\e=0.12$ and $\om=\frac{1}{2}$, using various function as indicated.}  \label{fig-vis2-e012-om1kroz2-variousfunctions}
\end{center}  \end{figure}

As discussed at length in~\cite{vis1}, the choice of function used for time averaging determines which details of the invariant sets will be revealed. The same holds for HTA: in order to capture all periodic set at given frequency, we need to consider more linearly independent functions. To illustrate this, in Fig.\ref{fig-vis2-e012-om1kroz2-variousfunctions} we show MHP for $\om=\frac{1}{2}$ at $\e=0.12$ using four different functions.
Each function identifies different phase space features belonging to the period-2 phase space partition. Two plots on the left isolate further period-2 structures within large-scale periodic sets. Two plots on the right fail to detect the large central period-2 set due to function being even in $x$. Chaotic zone is equally well pronounced in all plots.

\begin{figure}[!ht]  \begin{center} 
       \includegraphics[width=0.4\textwidth]{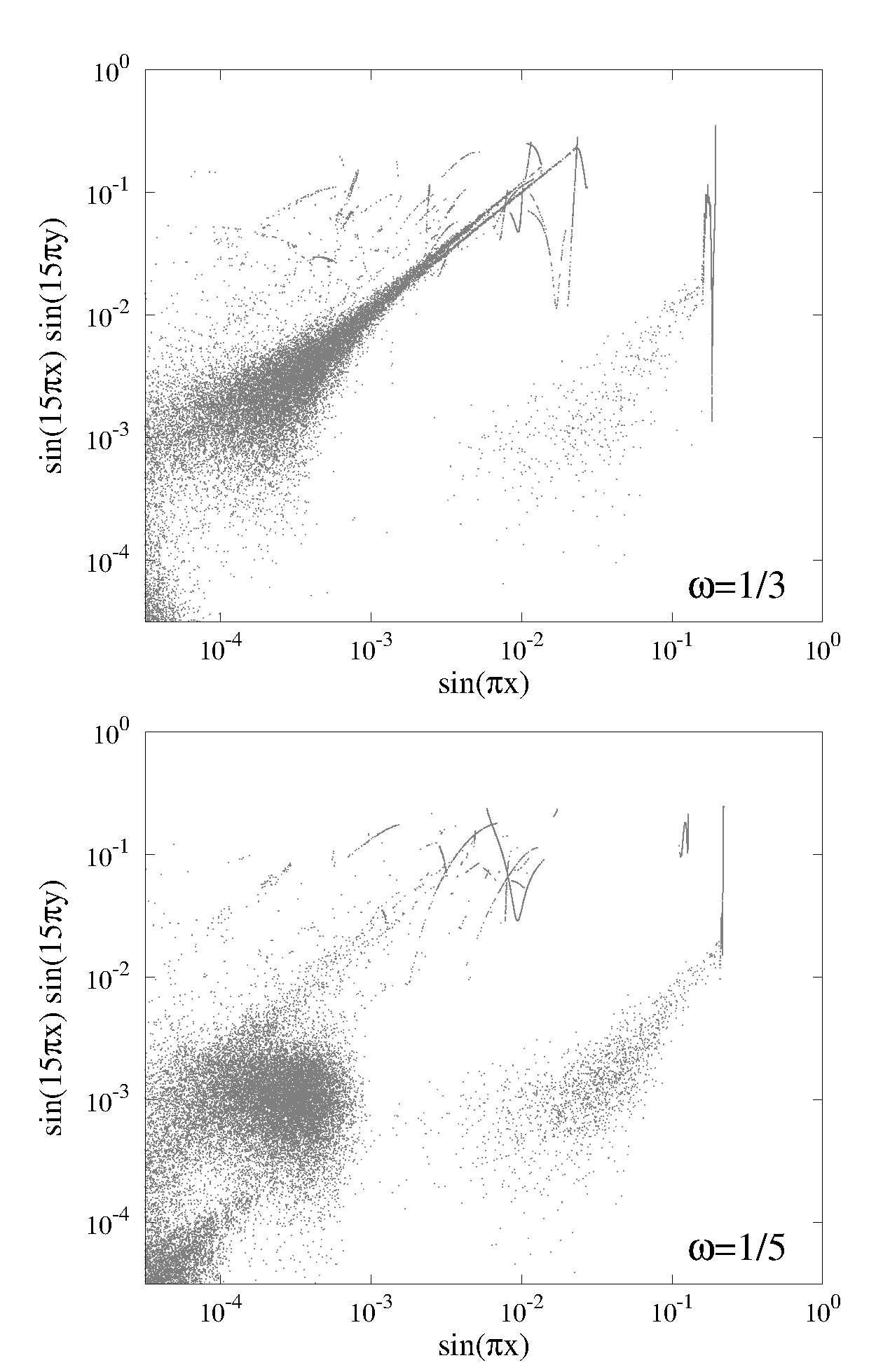}
  \caption{Scatter plots of two functions (indicated on axes) for $\e=0.12$ and $\om=\frac{1}{3}$, $\om=\frac{1}{5}$.}
 \label{fig-scatter-sin1pix-sin15pixsin15piy-e012-om1kroz2-om1kroz3}
\end{center}  \end{figure}

Given that each function identifies a potentially new periodic set within a given periodic partition, it is necessary to consider the information from multiple linearly independent functions. To illustrate this, we construct the Mesochronic Harmonic Scatter Plot (MHSP) by plotting the values of $h_\om$ for one function against the values for another function. In Fig.\ref{fig-scatter-sin1pix-sin15pixsin15piy-e012-om1kroz2-om1kroz3}, top and bottom, we show two scatter plots each done using two functions, with frequencies $\om=\frac{1}{3}$ and $\om=\frac{1}{5}$, respectively.
Each point in a scatter plot is defined by its $x$-coordinate ($h_\om$ for the first function) and by its $y$-coordinate ($h_\om$ for the second function). For both plots, the same pattern observed earlier are visible. The points with both coordinates in range of $O(10^{-4})$, $O(10^{-3})$ and $O(10^{-1})$, respectively correspond to non-resonating periodic set, chaotic zone and resonating periodic sets. Interestingly, between regions there are almost continuous transitions, which correspond to chaotic phase space points evolving in the vicinity of periodic sets. Note that now far more details on the resonating periodic sets are obtained, and could be used to visualize the resonating periodic partition more accurately.

\begin{figure}[!ht]  \begin{center} 
       \includegraphics[width=0.7\textwidth]{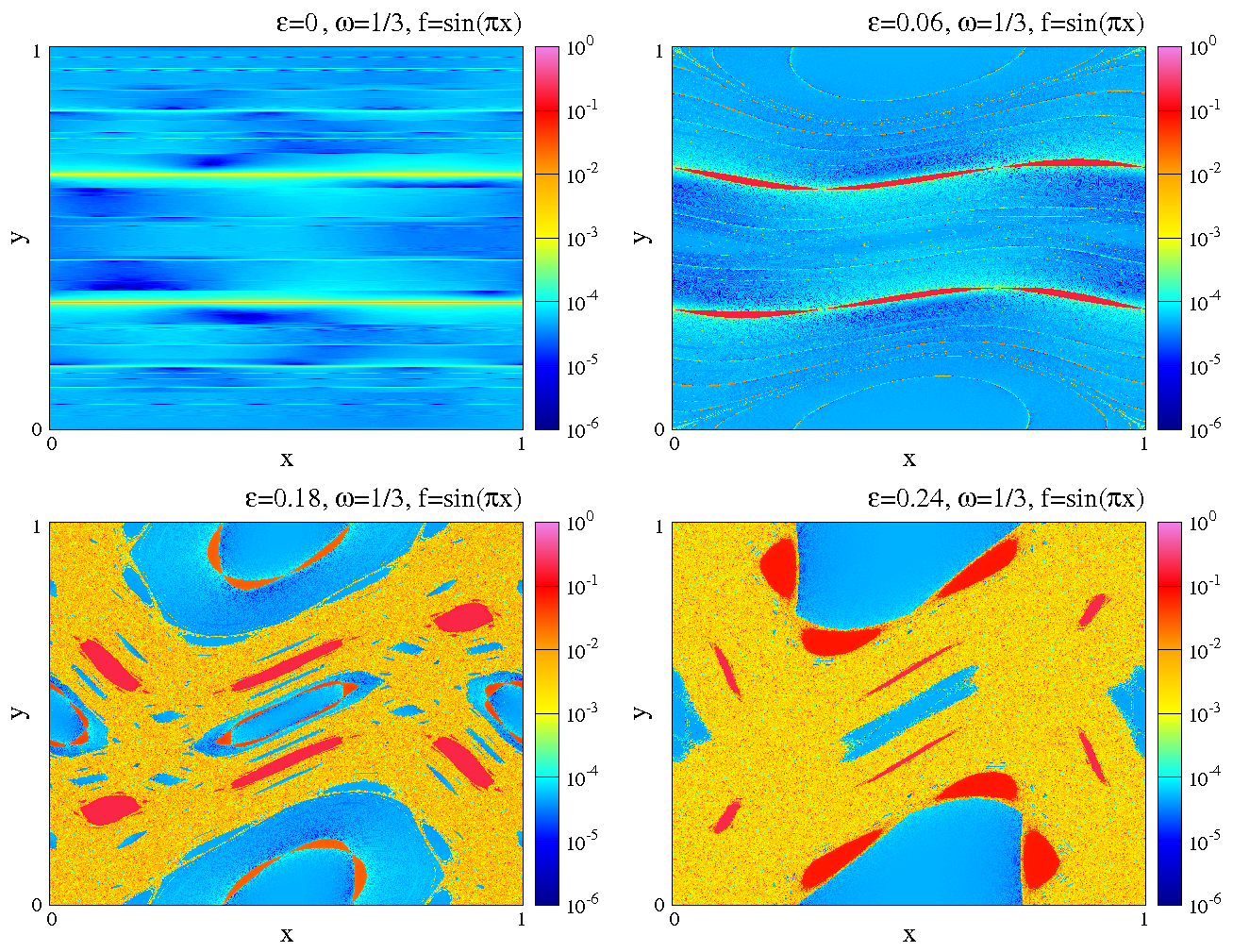}
  \caption{Single-function plots of $h_\omega$ for $\om=\frac{1}{3}$ and various values of $\e$.} \label{fig-vis2-sin1pix-om1kroz3-variouseps}
\end{center}  \end{figure}

Next we investigate the visualization of periodic set in relation to various values of $\e$ in the map Eq.\ref{eq-sm}. In Fig.\ref{fig-vis2-sin1pix-om1kroz3-variouseps} we show four MHPs for $\om=\frac{1}{3}$ and different $\e$-values. In the plot for $\e=0$, the lines $y=\frac{1}{3}$ and $y=\frac{2}{3}$ entirely consisting of period-3 orbits are clearly visible, in addition to $y=const$ lines consisting of higher periodicity orbits. 
The phase space regions near the periodic sets weakly resonate, since non-resonating trajectories whose structure is close to periodic, converge to zero more slowly, similarly to the chaotic zone. This effect is also present in plot for $\e=0.06$, where as expected, two main period-3 sets can be observed, in addition to many minor periodic sets of higher periodicity. Plot for $\e=0.18$ exhibits period-3 and period-6 sets of various shapes. Observe the secondary period-6 sets located around primary period-3 sets and central period-2 set. At this $\e$-value the chaotic zone can be clearly seen in a sharp contrast with both resonating and non-resonating periodic sets. Similar contrast is found for $\e=0.24$, where the chaotic zone is even more uniformly colored. 

\begin{figure}[!ht]  \begin{center} 
       \includegraphics[width=0.7\textwidth]{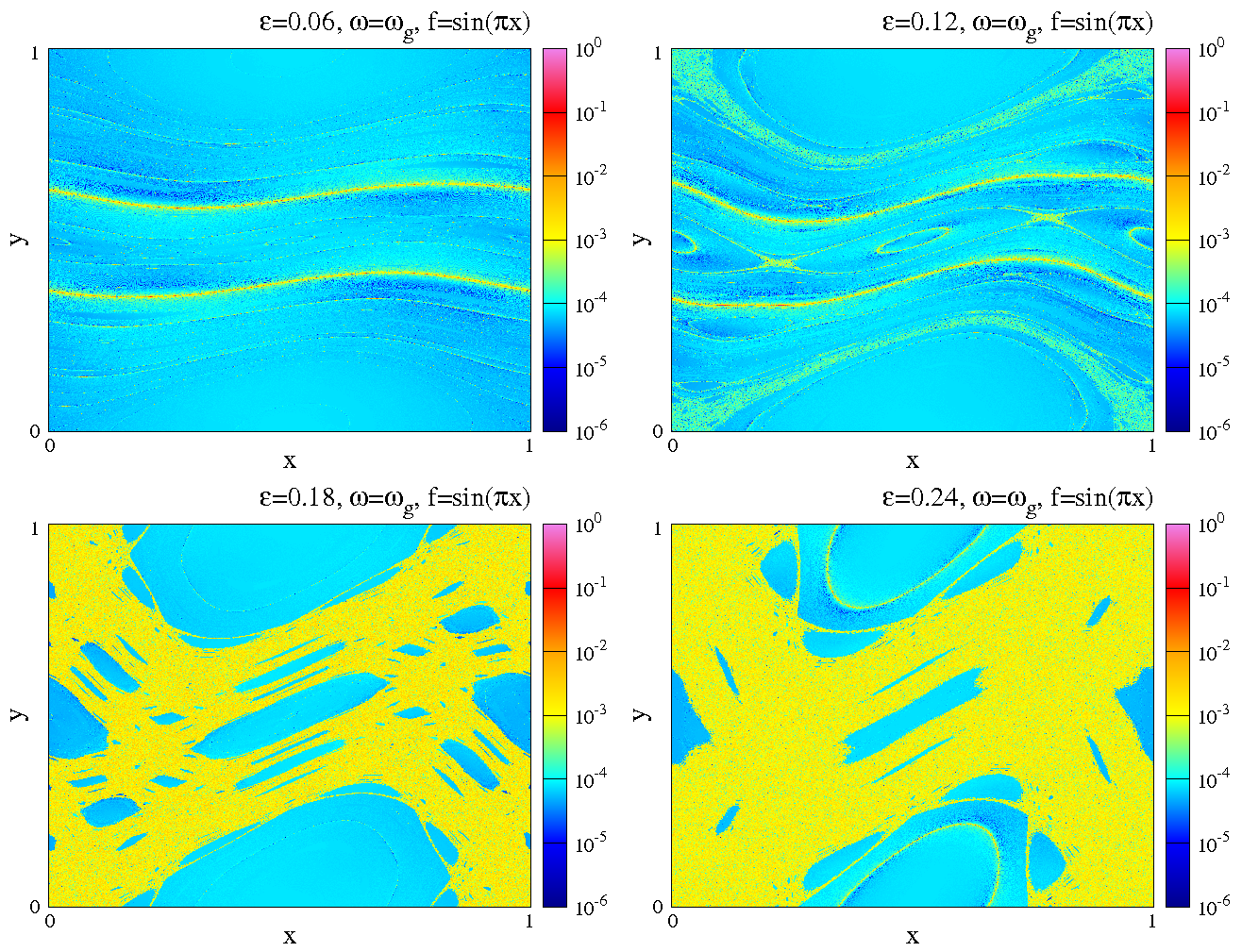}
  \caption{Single-function plots of $h_\om$ for irrational frequency $\om_g=\frac{\sqrt{5}-1}{2}$ and various values of $\e$.} \label{fig-vis2-sin1pix-omgolden-variouseps}
\end{center}  \end{figure}

Selection of rational frequencies up to $\frac{1}{2}$ allows identification of periodic sets with integer periodicities (note that choosing e.g. $\om=\frac{2}{5}$ still yields period-5 sets). However, standard map possesses an even richer variety of quasi-periodic orbits with irrational rotation numbers, which are expected to resonate at irrational frequencies. Chaotic zone is expected to resonate at all frequencies, including irrational ones. To examine this, we consider the golden mean frequency $\om_g = \frac{\sqrt{5}-1}{2}$. In Fig.\ref{fig-vis2-sin1pix-omgolden-variouseps} we show another four MHPs with increasing values of $\e$, computed for $\om=\om_g$. 
The invariant KAM curve for the corresponding frequency is clearly pronounced in $\e=0.06$ plot. In the plot for $\e=0.12$, the curve is still visible, together with the weakly resonating chaotic zone. Break-up of the golden mean KAM curve marks the chaotic transition in dynamical systems such as standard map~\cite{dynsys-books}. Our method can in principle be used for examining this transition. In the last two plots for $\e=0.18$ and $\e=0.24$ the chaotic zone is the only resonating phase space region, in addition to the secondary KAM golden mean curve located around elliptic fixed point $(\frac{1}{2},0)$. The resonating and non-resonating $h_\om$-values in all plots are again in the same value range as discussed above. The irrational frequency allows for easier visualization of the chaotic zone.

\begin{figure}[!ht]  \begin{center} 
       \includegraphics[width=0.5\textwidth]{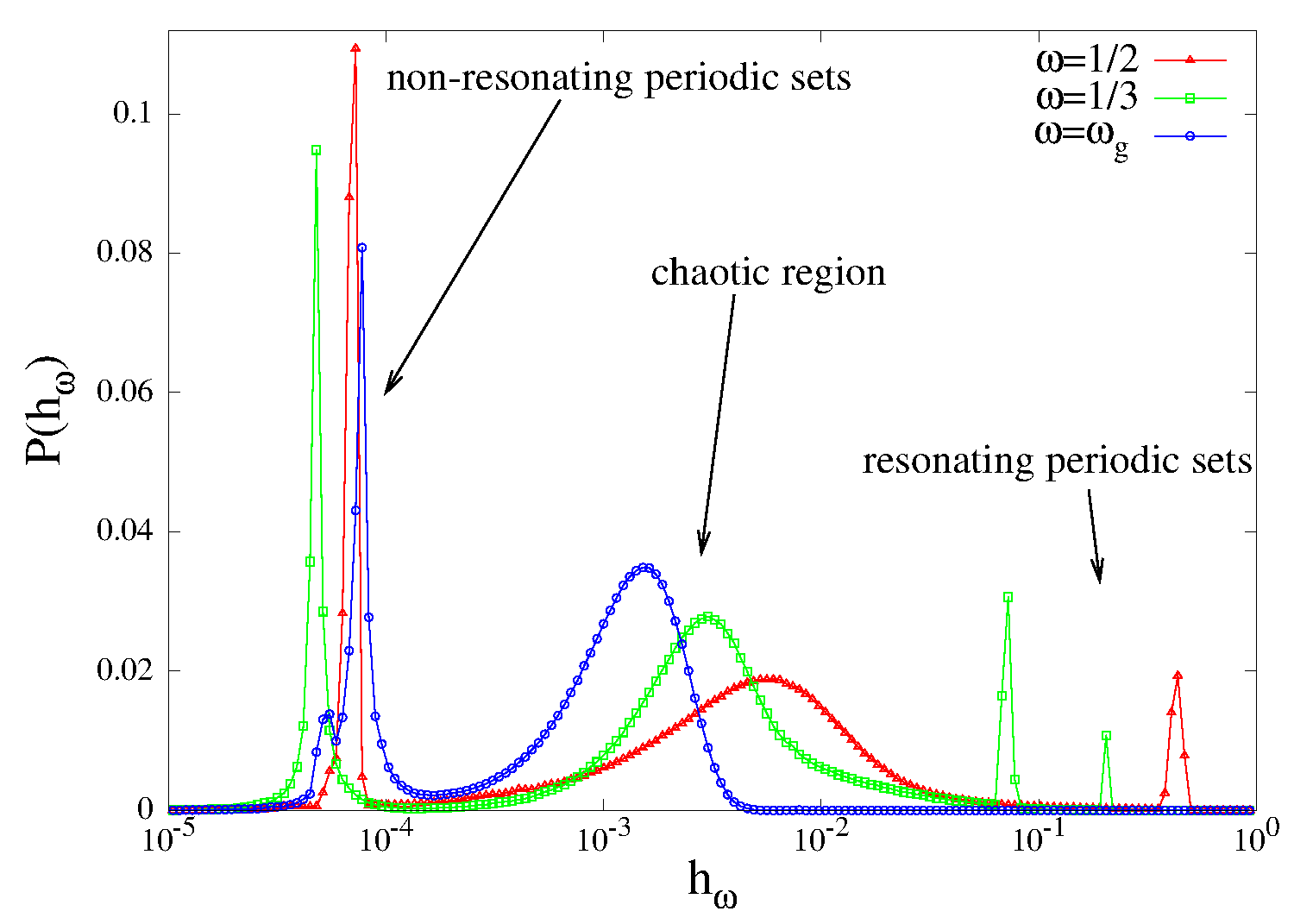}
  \caption{Distribution of $h_\om$-values for $\e=0.24$ and frequencies $\om = \frac{1}{2}, \frac{1}{3}, \om_g$. Peaks corresponding to different phase space regions are indicated.}
 \label{fig-vis2-hist-sin1pix-e024-variousom}
\end{center}  \end{figure}

We found that periodic and quasi-periodic sets can either resonate or not resonate, depending on the selected frequency. In contrast, the chaotic zone resonates \textit{for all frequencies}, although it always does so weakly. We show this quantitatively via histograms of $h_\om$-values, computed for different frequencies and reported in Fig.\ref{fig-vis2-hist-sin1pix-e024-variousom}. The parameter $\e=0.24$ is chosen so that the chaotic and regular regions occupy roughly similar portions of the phase space.
Histograms for $\om = \frac{1}{2}$ and $\om = \frac{1}{3}$ each consist of three well pronounced (groups of) peaks. Each corresponds to one phase space region, as indicated in the figure. The peaks are well separated; their distinction in fact grows with final iteration number $T$ (see next section). While all the peaks for periodic sets are very sharp, the chaotic peak displays a more smeared shape. In the histogram for $\om=\om_g$ the resonating peaks are very faint, while the other two peaks lie in the range of the corresponding peaks for rational frequencies. Chaotic zone consists of trajectories with a broad Fourier spectrum, which is reflected in our findings. This indicates, that for any selection of frequency, HTA indeed divides the phase space into three disjoint regions: periodic sets resonating at the selected frequency $\om$, periodic sets not resonating at the frequency $\om$ (i.e., set of periodicity incommensurable with $\om$), and the chaotic region (resonating at all frequencies).

\begin{figure}[!ht]  \begin{center} 
       \includegraphics[width=0.3\textwidth]{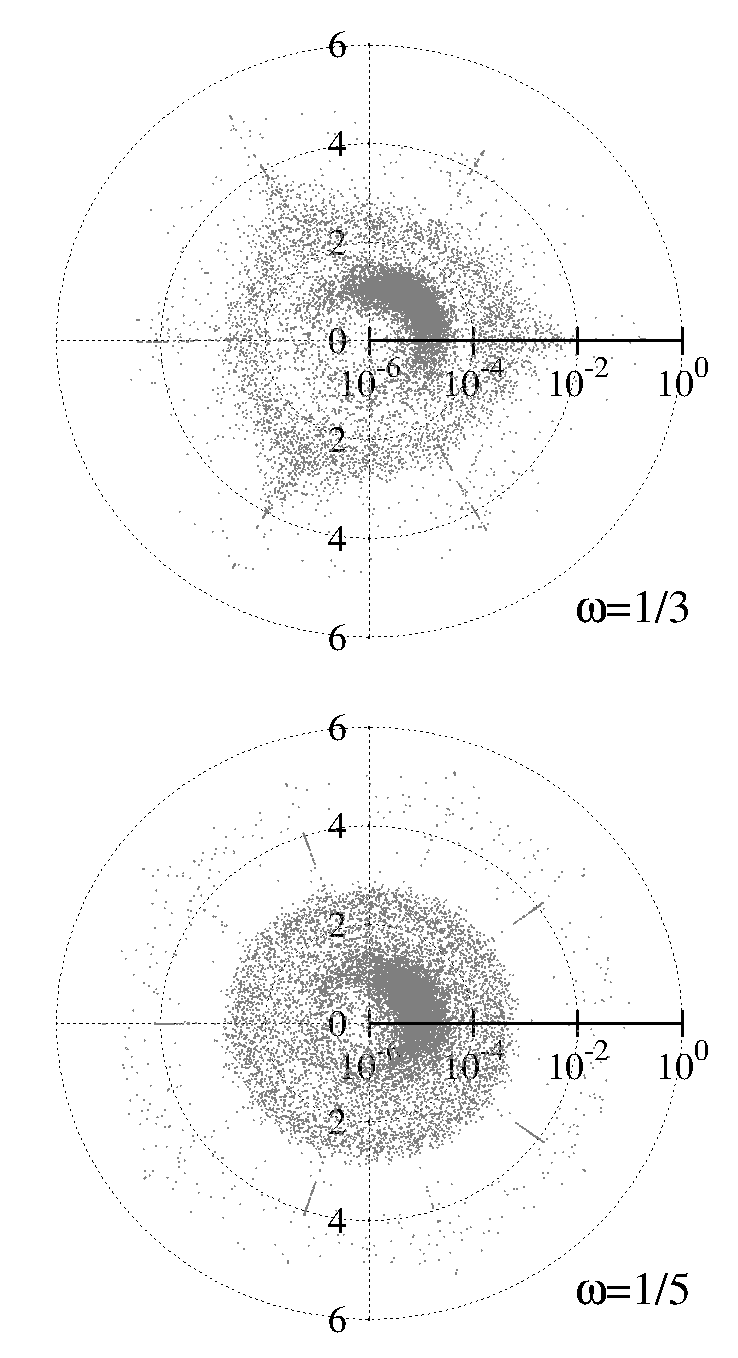}
    \caption{Snapshots of $f_\omega^*$-values in complex plane for $f(x,y)=\sin (\pi x)$, $\e=0.12$ and $\om=\frac{1}{3}, \frac{1}{5}$ (log-scale in radial coordinate). For better visibility, here we used $T=10^5$.}
\label{fig-vis2-polar-sin1pix-e012-variousom-time1e5}
\end{center}  \end{figure}

We finish this Section by studying the behavior of the complex phase in HTA. Recall that $f_\om^*$ is not an invariant function, since each iteration rotates it by the factor $e^{- i 2 \pi \omega}$. In Fig.\ref{fig-vis2-polar-sin1pix-e012-variousom-time1e5} we report the snapshots (not averages) of $f_\om^*$-values for all grid-points in the complex plane. Radial coordinate is in log-scale for consistency with other plots.
Along the radial coordinate, we immediately recognize three peaks considered in Fig.\ref{fig-vis2-hist-sin1pix-e024-variousom}: dense cluster around $|f_\om^*| \sim O(10^{-5})$ (non-resonating periodic sets), circular ring at $|f_\om^*| \sim O(10^{-3})$ (chaotic region), and outer points for $|f_\om^*| \sim O(10^{-1})$ (resonating periodic sets). The phase of $f_\om^*$ in the chaotic zone is uniformly randomized over the circle. In contrast, resonating periodic sets are consistent in phases; in profile for $\om=\frac{1}{3}$ we find six radially coherent groups of $f_\om^*$-values symmetrically organized around the center, each group identifying a separate subset of period-3 and period-6 sets. Similarly, five radially coherent groups of $f_\om^*$-values are corresponding to the resonating period-5 sets in profile for $\om=\frac{1}{5}$, along with a rim of scattered points representing smaller sets of higher periodicity. The complex phase of HTA therefore reflects the periodicity of the phase space region.

\begin{figure}[!ht]  \begin{center} 
       \includegraphics[width=0.7\textwidth]{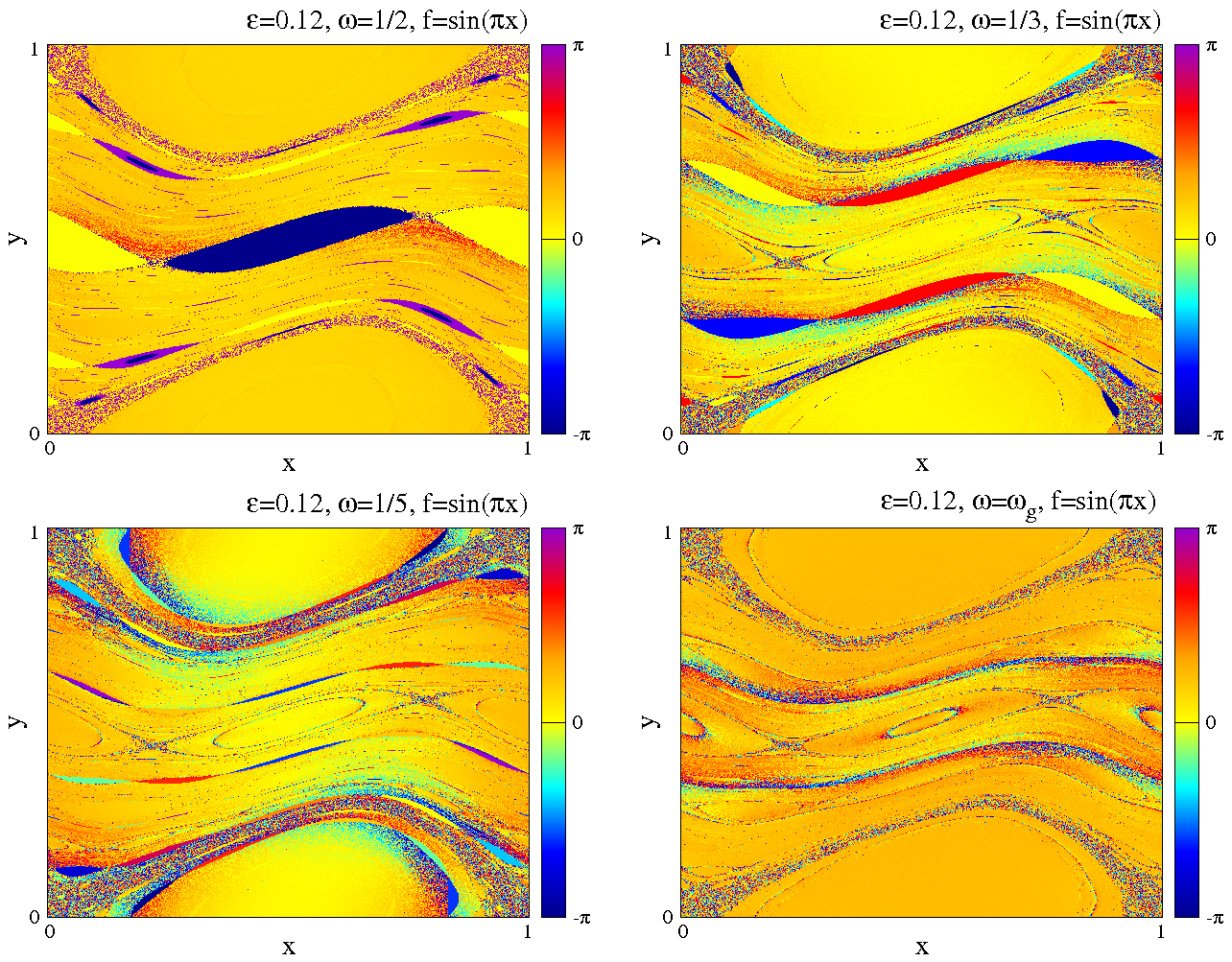}
  \caption{Snapshots of HTA phase values $\arg f_\omega^*$ taken at $T=30000$ iterations (not MHPs), for $\e=0.12$ and various frequencies $\om$.}
\label{fig-vis2-sin1pix-e012-variousom-phi}
\end{center}  \end{figure}

To illustrate the behavior of complex phase in the dynamical space, we show in Fig.\ref{fig-vis2-sin1pix-e012-variousom-phi} four snapshots of phases $\arg f_\omega^*$ (not MHPs), pictured through the cyclic colorbar. 
As already observed in Fig.\ref{fig-vis2-polar-sin1pix-e012-variousom-time1e5}, all resonating periodic sets exhibit coherent phase values, symmetrically organized among the subsets. In $\om=\frac{1}{3}$ plot for instance, three subsets of large period-3 sets have respectively $\arg f_\omega^* \simeq 0$, $\arg f_\omega^* \simeq \frac{\pi}{3}$ and $\arg f_\omega^* \simeq -\frac{\pi}{3}$. Similarly, each subset of period-5 sets in $\om=\frac{1}{5}$ plot has a defined phase value $\pi \pm \frac{2\pi n}{5}$ for $n=0,1,2$. On the other hand, chaotic zone exhibits a completely randomized phase values, uniformly distributed in $[-\pi,\pi)$, which is another way to discern it from the regular region. Non-resonating periodic sets are relatively uniform in phase value.

%% --------------------------------------------------------------------------------------------------------------------
%% --------------------------------------------------------------------------------------------------------------------
%% --------------------------------------------------------------------------------------------------------------------

\section{The Convergence Properties} \label{The Convergence Properties}

In this Section we examine how $h_\om$ converge to their limit values. As we show, the convergence regime directly depends on the dynamical nature of the considered trajectory, and can thus be used to establish the number of iterations $T$. 

\begin{figure}[!ht]  \begin{center} 
       \includegraphics[width=0.5\textwidth]{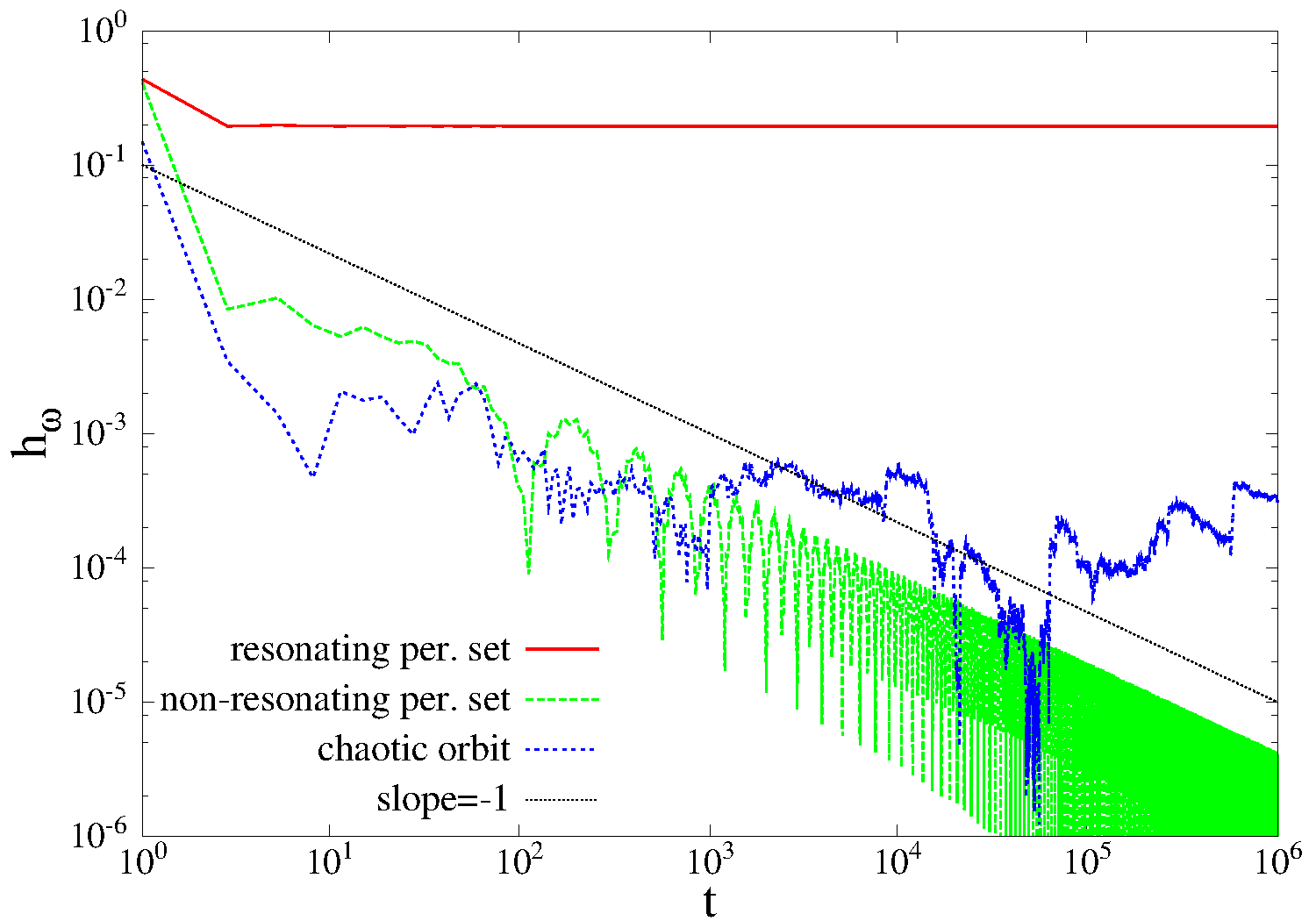}
    \caption{The value of $h_\om$ as function of time, for $\e=0.12$ and $\om=\frac{1}{3}$, and for three different initial points, belonging to different dynamical regions.} 
\label{fig-conv-sin1pix-e012-om1kroz3}
\end{center}  \end{figure}

We take $f(x,y)=\sin (\pi x)$ for $\om=\frac{1}{3}$ at $\e=0.12$, and consider the evolution of $h_\om$ as function of time. In Fig.\ref{fig-conv-sin1pix-e012-om1kroz3} we show the value of $h_\om$ obtained after $t$ iterations, for three different points representing different dynamical regions.
While the convergence of $h_\om$ is fast for both resonating and non-resonating phase space point, it becomes very irregular for the chaotic point. In fact, the convergence properties are equivalent to those for the usual time averages~\cite{vis1}. For regular trajectories the convergence rate is given by $\frac{const.}{t}$, which refers to both resonating and non-resonating periodic sets. In chaotic region, the convergence is much less predictable; the error can in general be bounded from below by $\frac{1}{\sqrt{t}}$. This also depends on the strength of chaotic motion -- for very strong chaos (mixing), the convergence rate approaches $\frac{1}{\sqrt{t}}$. These properties remain the same if the irrational frequencies are considered: for quasi-periodic orbits, whether resonating or not, the convergence exhibits $\frac{const.}{t}$ rate. In contrast, for chaotic zone, the convergence is again at best given by $\frac{1}{\sqrt{t}}$.

\begin{figure}[!ht]  \begin{center} 
       \includegraphics[width=0.5\textwidth]{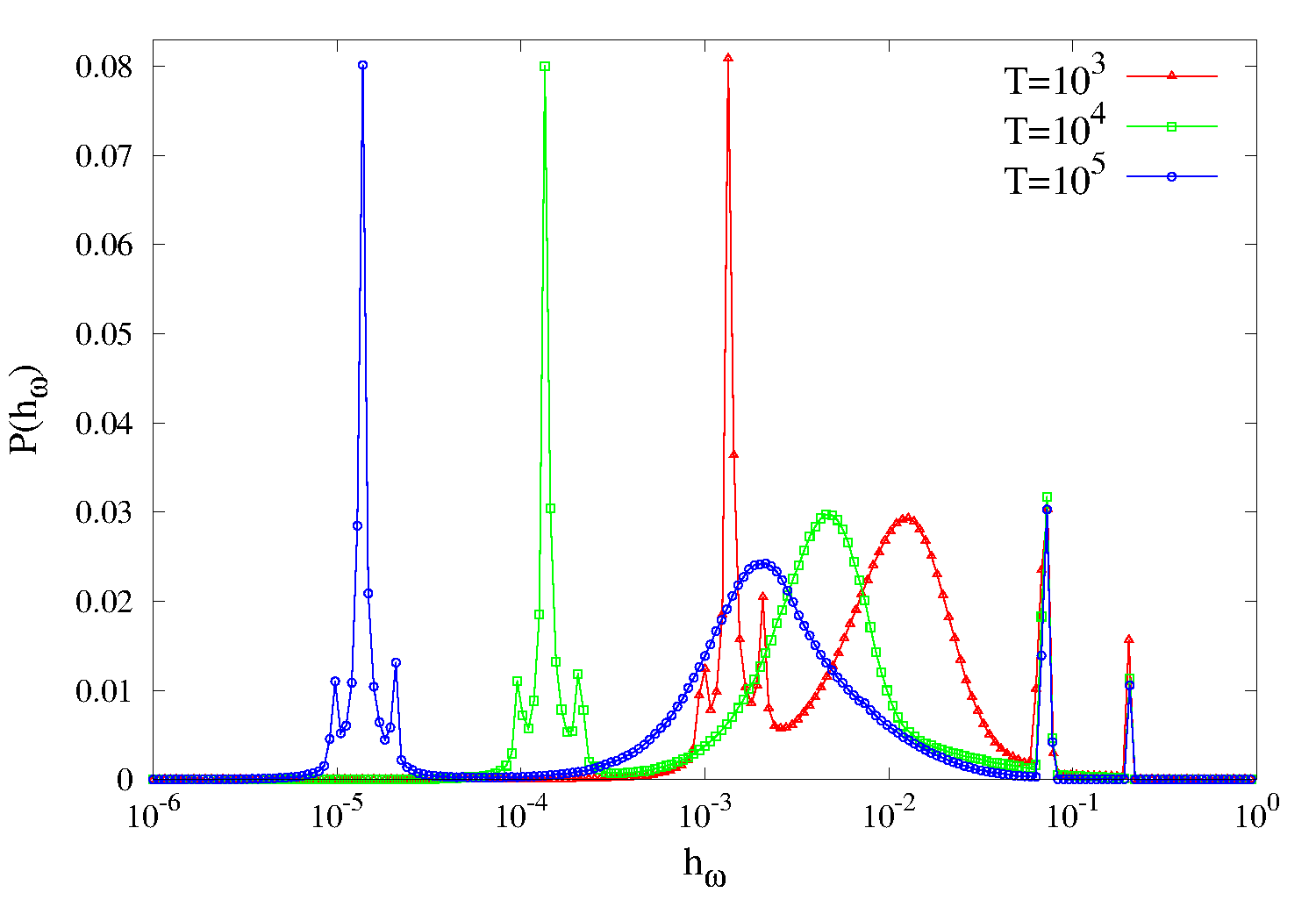}
  \caption{Distribution of $h_\om$-values for $\e=0.24$ and $\om=\frac{1}{3}$ (log-scale on $x$-coordinate), for three different final times $T$.}
\label{fig-hist-sin1pix-e024-om1kroz3-timeevol}
\end{center}  \end{figure}

We now examine the applicability of our method in detecting the chaotic zone. In Fig.\ref{fig-hist-sin1pix-e024-om1kroz3-timeevol} we show three histograms of $h_\om$ values, obtained for three different values of $T$. Three groups of peaks, corresponding to resonating, chaotic, and non-resonating phase space parts can be recognized in all three histograms (cf. Fig.\ref{fig-vis2-hist-sin1pix-e024-variousom}). 
The resonating peaks do not move. In contrast, both chaotic and non-resonating peaks move towards zero, but with different speeds. The speed of the non-resonating peaks is roughly double the speed of the chaotic peaks. This proves that the convergence rate for the chaotic zone is on average $\frac{1}{\sqrt{t}}$, meaning that for a sufficiently long time it can be clearly distinguished from the non-resonating region. Note also that the shape of peaks does not change significantly as they move. Fig.\ref{fig-hist-sin1pix-e024-om1kroz3-timeevol} considers $\e=0.24$ where the chaos is strong; while this result does hold for a general $\e$-value, the separation of peaks might be slower for smaller $\e$ values, or might be less clear due to smaller chaotic zone.

%% --------------------------------------------------------------------------------------------------------------------
%% --------------------------------------------------------------------------------------------------------------------
%% --------------------------------------------------------------------------------------------------------------------

\section{The Froeschl\'e Map} \label{The Froeschle Map}

In the previous Sections we illustrated the applicability of our visualization method relying on well-known standard map. In the following two Sections, we demonstrate the method's full potential by studying more complicated higher-dimensional dynamical systems. 

We begin with the Froeschl\'e map, which is a measure-preserving mapping of 4-dimensional cube $[0,1]^4$ onto itself, introduced in the context of celestial dynamics~\cite{froeschle}. It is defined as:
\begin{equation} \begin{array}{llc}
x_{1}' &= \; x_{1} + y_{1}  + \e_{1} \sin (2 \pi x_{1}) + \eta \sin (2 \pi x_{1} + 2 \pi x_{2})      \;\;\;  &[mod \; 1]   \\
y_{1}' &= \; y_{1} + \e_{1} \sin (2 \pi x_{1}) + \eta  \sin (2 \pi x_{1} + 2 \pi x_{2})              \;\;\;  &[mod \; 1]   \\
x_{2}' &= \; x_{2}  + y_{2} + \e_{2}  \sin (2 \pi x_{2}) + \eta  \sin (2 \pi x_{1} + 2 \pi x_{2})    \;\;\;  &[mod \; 1]  \\
y_{2}' &= \; y_{2} + \e_{2} \sin (2 \pi x_{2}) + \eta  \sin (2 \pi x_{1} + 2 \pi x_{2})              \;\;\;  &[mod \; 1] 
\end{array} \label{froeschlemap} \end{equation}
and it can be seen as two symplecticly coupled standard maps. For simplicity, we set $\e_1=\e_2=\e$ and focus on the scenario of two identical standard maps, interacting via coupling parameter $\eta$. Following the ergodic partition analysis of this map from~\cite{vis1}, we here extend our study to MHPs.

\begin{figure}[!ht]  \begin{center} 
       \includegraphics[width=\textwidth]{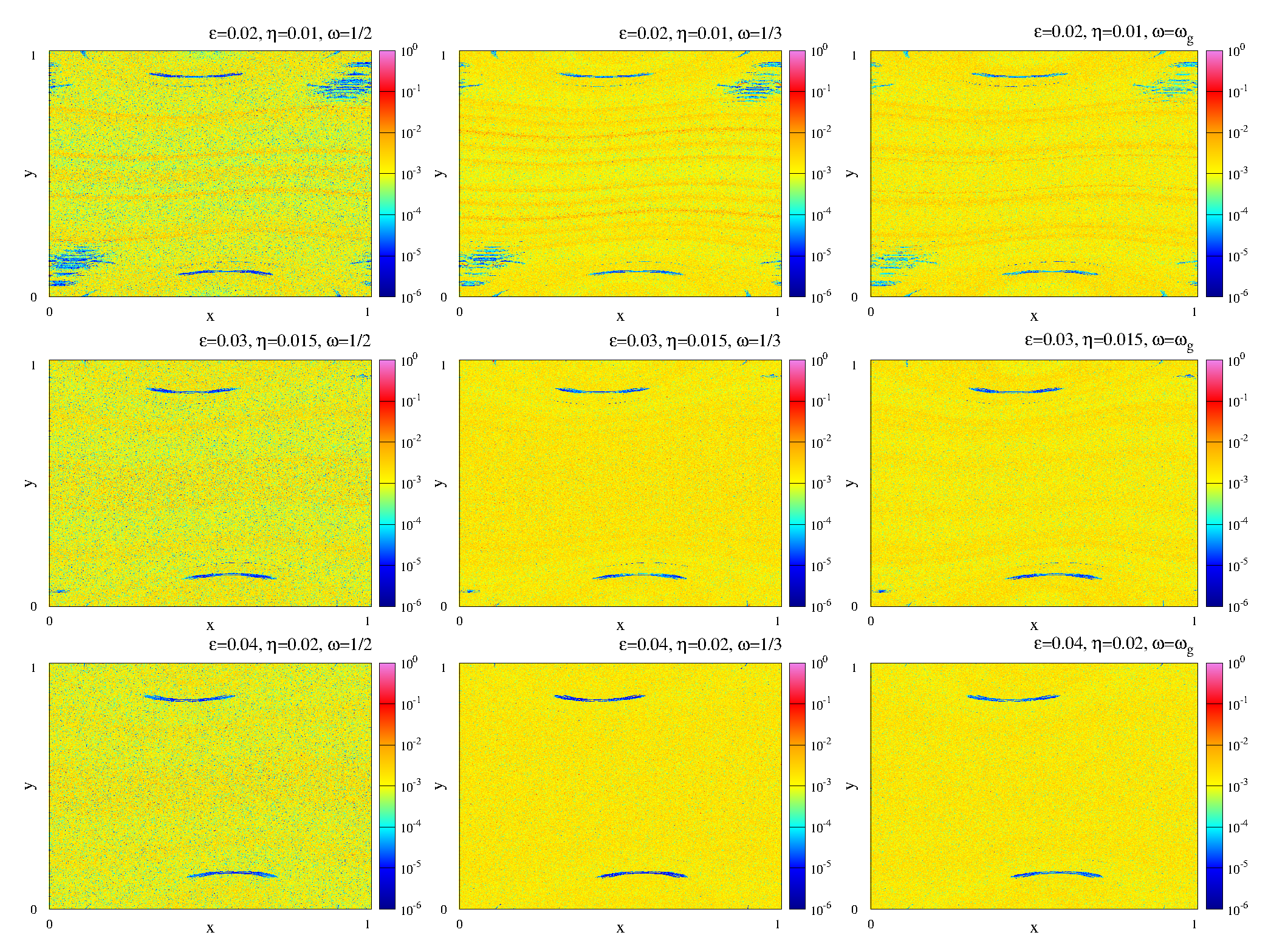}
  \caption{MHPs of function $f(x_1,y_1,x_2,y_2) = 2 \cos (2 \pi y_1) + \cos (2 \pi y_1) \cos (2 \pi y_2) + \cos (12 \pi x_2) + \cos (12 \pi x_1)$ under the dynamics of Froeschl\'e map for $\om=\frac{1}{2},\frac{1}{3},\om_g$, computed on the phase space section $(x_2=0,y_2=0)$ over the grid of $500 \times 500$ points. The dynamics was run for $T=2 \times 10^5$ iterations. The values of $\e$ and $\eta$ are linked by $\e=2\eta$ and indicated in each plot.} 
\label{fig-vis2-fro-f-e002003004-T2e5-om1kroz23golden}
\end{center}  \end{figure}

An elegant way to simplify the parameters in Froeschl\'e map without losing its dynamical richness it to fix $\e=2\eta$. This reduces our analysis to variations of a single parameter. First we show MHPs of 2D slice (2D section in 4D phase space). To that end we construct a grid of $500 \times 500$ points in $(x_1,y_1)$-space on the section given by $(x_2,y_2)=(0,0)$. We consider the function $f(x_1,y_1,x_2,y_2) =  2 \cos (2 \pi y_1) + \cos (2 \pi y_1) \cos (2 \pi y_2) + \cos (12 \pi x_2) + \cos (12 \pi x_1)$ and run the dynamics for $2 \times 10^5$ iterations. Three frequencies are considered: rational $\om=\frac{1}{2}$ and $\frac{1}{3}$, and irrational $\om_g$. For each frequency, three values of parameters $\e$ and $\eta$ are considered. The resulting MHPs are shown in Fig.\ref{fig-vis2-fro-f-e002003004-T2e5-om1kroz23golden}. 
Parts of periodic sets are visible in each plot as either resonating or non-resonating, together with the weakly resonating chaotic zone. As expected, the chaotic zone grows with the increase of the coupling parameters, while the periodic region becomes gradually smaller. Remember that in general, invariant sets are here four dimensional, meaning that only their 2D projections can be visible in these section plots. This explains why the number of subsets does not match their periodicity. Also, note that the periodic sets are in fact mostly visible as non-resonating. This means they are of periodicities incommensurable with the considered periodicities (frequencies). This illustrates how our method can be employed to detect and visualize periodic sets of unknown periodicity, although without actually revealing the periodicity itself.

\begin{figure}[!ht]  \begin{center} 
       \includegraphics[width=0.9\textwidth]{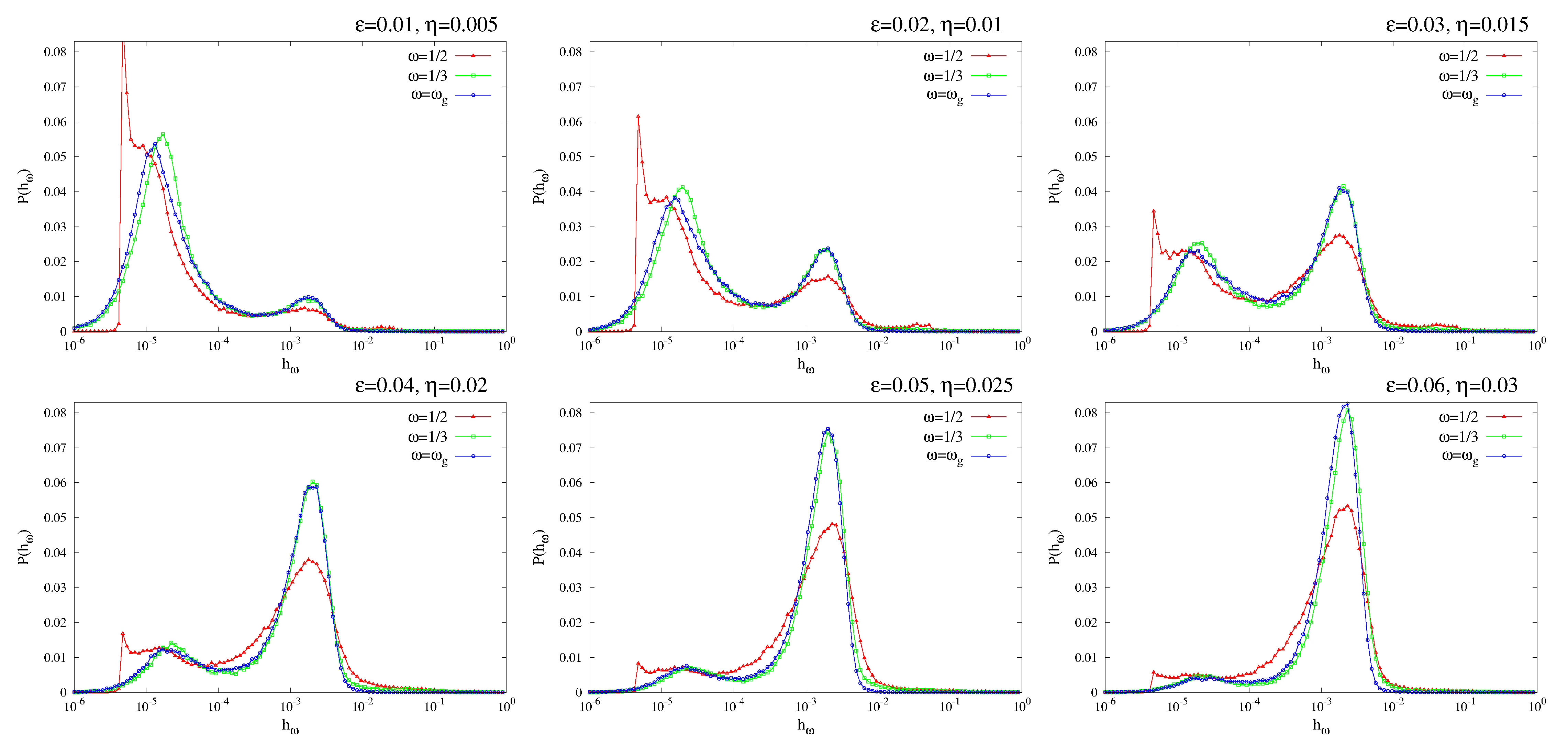}
  \caption{Histograms of $h_\om$ values computed for $f(x_1,y_1,x_2,y_2) = 2 \cos (2 \pi y_1) + \cos (2 \pi y_1) \cos (2 \pi y_2) + \cos (12 \pi x_2) + \cos (12 \pi x_1)$ under the dynamics of Froeschl\'e map for $\om=\frac{1}{2},\frac{1}{3},\om_g$ for a 4D grid of $18 \times 18 \times 18 \times 18$ random phase space points. The dynamics was run for $T=2 \times 10^5$ iterations. The values of $\e$ and $\eta$ are linked by $\e=2\eta$ and indicated in each histogram.} 
\label{fig-vis2-hist-fro-e00123456-4D-T2e5-om1kroz23golden}
\end{center}  \end{figure}

Next we examine the global evolution of the periodic sets with variations of $\e=2\eta$. We start with $18 \times 18 \times 18 \times 18$ phase space points, which we pick at random from $[0,1]^4$ in order to avoid hitting too many resonances. For each initial point we compute $h_\om$ after $T=2 \times 10^5$ iterations for the same function as in Fig.\ref{fig-vis2-fro-f-e002003004-T2e5-om1kroz23golden}, and for the same three frequencies. For each set of $\e,\eta$ values, the histogram of $h_\om$ values is computed, and the results are shown in Fig.\ref{fig-vis2-hist-fro-e00123456-4D-T2e5-om1kroz23golden}.
Each histogram plot consists of two groups of peaks, one corresponding to the chaotic region (near $10^{-3}$) and the other to non-resonating periodic sets (near $10^{-5}$). As already noted, periodic sets in the phase space of Froeschl\'e map are in general of high periodicities, incommensurable with the three considered frequencies. Despite not observing periodic sets of substantial size, we do know that many periodic sets actually exit, which is suggested by the large volume of phase space points clearly displaying non-resonating values of $h_\om$. With increase of parameters $\e,\eta$, we observe the transition to full phase space chaos, as indicated by the growth of the chaotic peak. As discussed in~\cite{vis1}, the chaotic transition for Froeschl\'e map is qualitatively similar to that of standard map (break-up of invariant KAM tori), although it appears to involve more complicated mechanisms. In agreement with what observed there, we here conclude that the global merging of resonances occurs around the value of $\e = 2\eta \simeq 0.05$.

%% --------------------------------------------------------------------------------------------------------------------
%% --------------------------------------------------------------------------------------------------------------------
%% --------------------------------------------------------------------------------------------------------------------

\section{Extended Standard Map} \label{Extended Standard Map}  

In this Section we study the extended standard map using HTA. The map has been proposed in~\cite{esm} as a three-dimensional generalization of standard map. It is a measure-preserving action-action-angle mapping of the cube $[0,1]^3$ onto itself, defined by:
\begin{equation} \begin{array}{lllc}
x' &= x + \e \sin (2\pi z) + \delta \sin (2\pi y)      \;\;\;  &[mod \; 1]  \\
y' &= y + \e \sin (2\pi z)                             \;\;\;  &[mod \; 1]  \\  
z' &= z + x + \e \sin (2\pi z) + \delta \sin (2\pi y)  \;\;\;  &[mod \; 1]
\end{array} \label{esm} \end{equation} 
When either $\e$ or $\delta$ are zero, the phase space can be divided into parallel 2D planes, such that dynamics on each of plane is isolated from the dynamics on other planes. On the other hand, when both $\e$ and $\delta$ are non-zero, this dynamical system is argued to be ergodic over the entire phase space~\cite{esm}. In fact, no invariant surface persists once both perturbations are simultaneously positive. On top of the numerical evidence already provided via ergodic partitioning in~\cite{vis1}, we here provide an additional argument supporting this claim.

\begin{figure}[!ht]  \begin{center} 
       \includegraphics[width=0.5\textwidth]{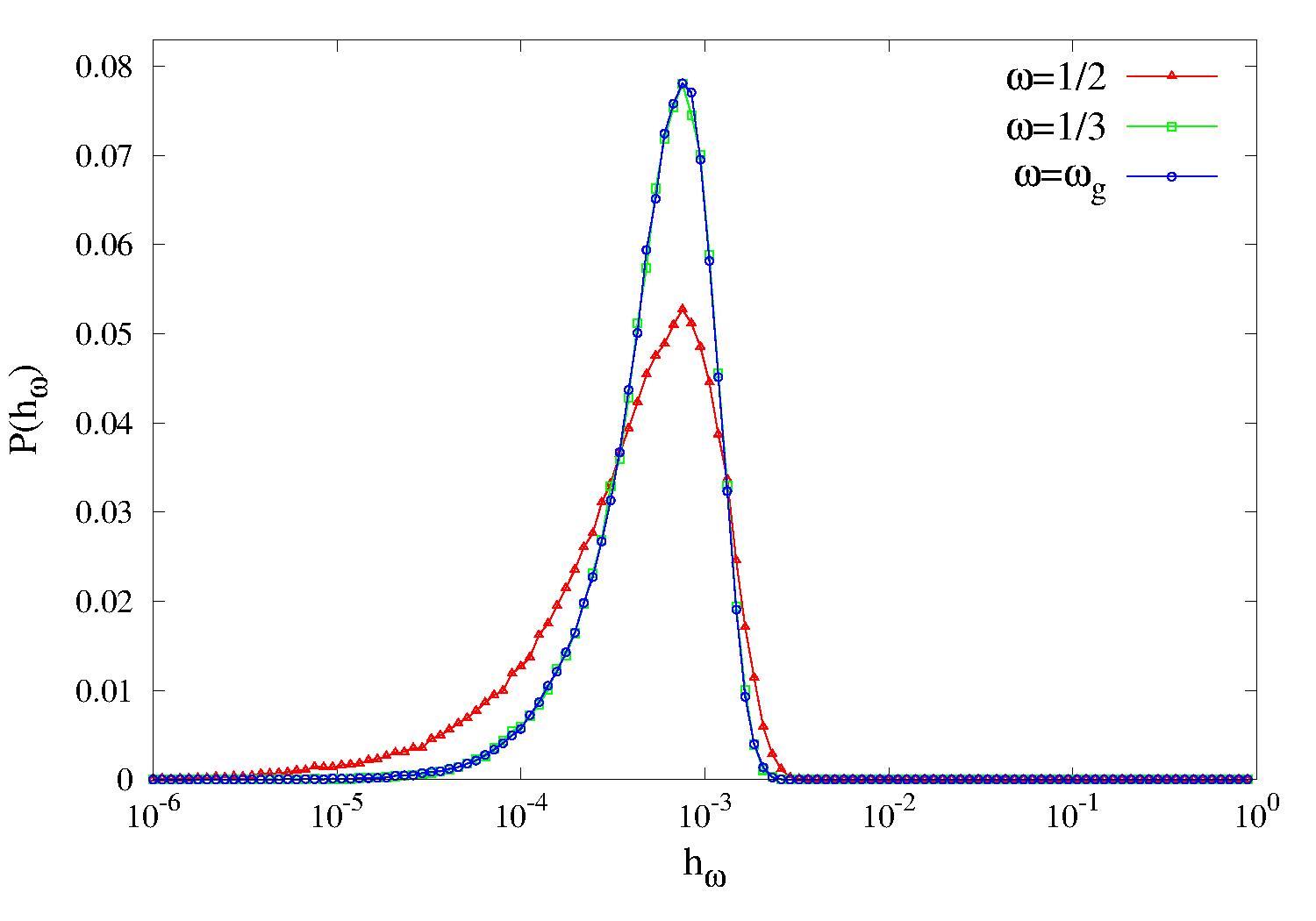}
  \caption{Distribution of $h_\om$-values for extended standard map with $\e=0.01$ and $\delta=0.001$ for a 3D grid of $50 \times 50 \times 50$ points covering the phase space. $h_\om$ for $f(x,y,z) = \sin(4\pi y) \cos(4\pi x) \cos(12\pi z)$ are computed for $T=2 \times 10^5$ iterations, with frequencies as indicated.}
\label{fig-hist-esm-e001del0001-3D-T2e5-om1kroz23golden}
\end{center}  \end{figure}

We set a three-dimensional grid of $50 \times 50 \times 50$ points covering the phase space of extended standard map with $\e=0.01$ and $\delta=0.001$. We run the HTA of function $f(x,y,z) = \sin(4\pi y) \cos(4\pi x) \cos(12\pi z)$ for $T=2 \times 10^5$ iterations, and for three frequencies $\om=\frac{1}{2},\frac{1}{3},\om_g$. We show the histograms of the obtained values of $h_\om$ in Fig.\ref{fig-hist-esm-e001del0001-3D-T2e5-om1kroz23golden}. 
For all three histograms, both the range of values and the bell-shape of the curves, indicate that all grid points lie in the chaotic region of the phase space. This suggests that there are no periodic sets of the considered periodicities of any size. In fact, it implies that there are no periodic sets of any periodicity, since any periodic set of frequency different from the considered would be visible as non-resonating. Note that irrational frequency $\om_g$ agrees with this result. This clearly confirms the claim of the extended standard map being ergodic. This point also serves to note the usefulness of our method: we employed HTA analysis to study higher-dimensional dynamical systems, much more complicated than the standard map.

%% --------------------------------------------------------------------------------------------------------------------
%% --------------------------------------------------------------------------------------------------------------------
%% --------------------------------------------------------------------------------------------------------------------

\section{Discussion and Conclusions} \label{Discussion and Conclusions}

We exposed a new method of analysis of measure-preserving dynamical systems based on harmonic time averages (HTA) proposed in~\cite{mb}, with roots in both frequency analysis and ergodic theory, as a continuation of our earlier work on ergodic partition theory~\cite{vis1,mezic}. Both our methods are realized as computational algorithms for detection and visualization of invariant and periodic sets in the phase space of arbitrary measure-preserving dynamical system. We showed the method's implementation using the known properties of Chirikov standard map. Of course, our method is in no way limited to discrete-time dynamical systems (maps) considered here. Except for bigger computational demand, the method is fully implementable for continuous-time dynamical systems. The distribution of values of $h_\om$ was analyzed, resulting in the identification of histogram peaks that correspond to different dynamical regions, in particular, resonating periodic set, non-resonating periodic sets and chaotic region. The convergence issues have been considered in relation to estimating the precision of the obtained $h_\om$. The full potential of the method was demonstrated in an application to higher-dimensional dynamical systems, 4D Froeschl\'e map and 3D extended standard map. Known dynamical features of these systems were readily found, in addition to certain novel insights. 

Our method is applicable regardless of the system's integrability or other dynamical properties. The key generalization to follow, is the development of clustering algorithms, in analogy with MSPs examined in~\cite{vis1}, aimed at visualizing the entire periodic partition for a given frequency~\cite{romesburg,budisic1}. This is done by considering HTA of multiple functions with the same frequency and optimally clustering the scatter plot points (cf. Fig.\ref{fig-scatter-sin1pix-sin15pixsin15piy-e012-om1kroz2-om1kroz3}). The periodic partition is obtained upon identifying the phase space part that corresponds to each cluster of scatter plot points. When a sufficient number of linearly independent functions is considered, this clustering algorithm will reveal the periodic partition with any prescribed precision. Of course, this opens the question of optimal clustering, which may depend on the dynamical system and the corresponding scatter plot (e.g. shape/volume of the clustering cells). Furthermore, the relationships between the convergence properties and the nature of the underlying trajectory is to be further investigated. While the convergence slope gives a rough estimate of the type of the orbit, a detailed analysis separating between rational and irrational frequencies might yield further insights.

The full applicative extent of our method lies in high-dimensional systems, which can be investigated in various ways, for instance via MHPs done over 2D phase space sections. In fact, the method can be easily extended to distributed systems, such as complex networks~\cite{omer}, and in particular networks that model extended dynamical systems~\cite{ja2,ja1}. This could reveal novel details related to the collective dynamical behavior, usually displayed by such systems.  \\[0.8cm]

%% --------------------------------------------------------------------------------------------------------------------
%% --------------------------------------------------------------------------------------------------------------------
%% --------------------------------------------------------------------------------------------------------------------

\n {\bf Acknowledgments.} Work supported by the Republic of Slovenia via Creative Core FISNM-3330-13-500033, in addition to the ARRS program numbers P1-0383 and P1-0044, and project number J1-5454. Work also supported by the AFOSR grant numbers F49620-03-1-0096 and FA9550-09-1-0141. Part of this work was done during ZL's stay at University of California Santa Barbara, and part during his stay at Dept. of Theoretical Physics, Jo\v zef Stefan Institute, Ljubljana, Slovenia. We thank the colleagues Umesh Vaidya, Arkady Pikovsky, Dima Shepelyansky and Toma\v z Prosen for useful comments and discussions.

\begin{footnotesize}  \end{footnotesize}
\end{document}